%% file: iterators_arxiv.tex
\documentclass{article}

\usepackage[margin=1in]{geometry}

\usepackage{algorithm}
\usepackage{algorithmicx}
\usepackage{algpseudocode}
\usepackage{mdframed}
\usepackage{hyperref}
\usepackage{amsfonts}
\usepackage{float}
\usepackage{amssymb}

\usepackage{graphicx}
\usepackage{caption}
\usepackage{subcaption}
\usepackage{amsthm}

\usepackage{amsmath}
\usepackage{comment}
\usepackage{multicol}
\newtheorem{theorem}{Theorem}

\newtheorem{definition}{Definition}[section]

\newcommand{\Oh}{\mathcal{O}}

\newcommand{\ol}[1]{\overline{#1}}

\newcommand{\ins}{\textsc{Insert}}
\newcommand{\del}{\textsc{Delete}}
\newcommand{\con}{\textsc{Contains}}
\newcommand{\iter}{\textsc{Iterate}}

\newcommand{\dsins}{\textsc{DS-Insert}}
\newcommand{\dsdel}{\textsc{DS-Delete}}

\newcommand{\True}{\texttt{True}}
\newcommand{\False}{\texttt{False}}

\newcommand{\seek}{\textsc{Seek}}
\newcommand{\dsseek}{\textsc{DS-Seek}}

\newcommand{\clean}{\textsc{Cleanup}}

\newcommand{\etal}{{\em et~al.}}

\usepackage[english]{babel}   
\newcommand{\larr}{\ensuremath{\leftarrow}}
\newcommand{\tryReport}{\textsc{Try-Report}}
\renewcommand{\mark}{\textsc{Mark}}

\newcommand{\hnode}{\textsc{HNode}}
\newcommand{\fset}{\textsc{FSet}}

\definecolor{todocol}{RGB}{170,15,45}

\title{Linearizable Iterators for Concurrent Sets}

\author{Archita Agarwal \and Zhiyu Liu \and Eli Rosenthal \and Vikram Saraph}

\begin{document}


\maketitle
\begin{abstract}

This paper proposes a general framework for adding linearizable iterators to a class of data structures that implement set operations. We introduce a condition on set operations, called \emph{local consistency}, which informally states that set operations never make elements unreachable to a sequential iterator's traversal. We show that sets with locally consistent operations can be augmented with a linearizable iterator via the framework. Our technique is broadly applicable to a variety of data structures, including hash tables and binary search trees. We apply the technique to sets taken from existing literature, prove their operations are locally consistent, and demonstrate that iterators do not significantly affect the performance of concurrent set operations.

\end{abstract}


\section{Introduction}
\input{intro}

\subsection{Related work}
\input{related_work}

\section{Iterator framework}
\input{framework_arxiv}

\input{theory_arxiv}

\input{applications_appendix}

\section{Performance evaluation}
\input{results}

\bibliographystyle{plain}
\bibliography{iterators_submission_europar}



\end{document}

%% file: intro.tex
The increasing prevalence of distributed and parallel computer
architectures has motivated the development of a large selection of concurrent
data structures. It is desirable for these data structures to come equipped
with \emph{lock-free} operations, in order to guarantee system-wide progress
and a certain amount of resilience to failure. Furthermore,
\emph{linearizability} \cite{Herlihy90} is often required
to ensure consistency among the data structure's operations, so
that any execution appears to take place one operation at a time. However,
designing efficient data structures that are lock-free and linearizable can
be challenging, requiring techniques highly specific to the data structures in question.

In many application domains, one desires functionality that can retrieve
information dependent on the global view of a data structure. For example, one
may wish to calculate the size of a data structure at a moment in time, or to
apply a filter to all its elements. In a single-threaded
setting, one can easily accomplish these tasks through \emph{iteration}: a
common API that provides sequential access to \emph{collect} all elements
of a data structure. An \iter{} operation returns this sequence of elements, called a \emph{snapshot}.
In a multithreaded setting, an iterator is \emph{linearizable} if the operation appears
to other threads to occur instantaneously. In other words, a linearizable iterator is required to
return a snapshot containing elements that could have all been simultaneously in the set.

Harris's lock-free linked list \cite{Harris2001} is a widely used set implementation supporting linearizable
\ins{}, \del{}, and \con{} operations (collectively called \emph{set} operations).
Petrank and Timnat \cite{Petrank2013} designed a linearizable iteration scheme for
Harris's list, capable of supporting multiple concurrent lock-free iterators.
In their implementation, iterators work together
to build a snapshot via a shared \emph{snap-collector} object, with concurrent set
operations \emph{reporting} their modifications.
However, their work relies on specific aspects of the linked list,
and applies only to other data structures sharing those particular aspects.
We propose a more general solution that applies to a broad class
of data structures.

\subsection{Contributions}

In order for an iterator to be linearizable, its snapshot must satisfy two properties: it should (1) be consistent with concurrent updates to the set and (2) contain all elements not modified by concurrent updates. We call property (2) \emph{global consistency}. 
Intuitively, property (1) ensures that modified nodes are accounted for correctly, while property (2) ensures that unmodified nodes are accounted for correctly.
Petrank and Timnat's snap-collector provides an iterator for a linked list implementation, whose snapshot is consistent with concurrent updates (Property 1). Furthermore, their linked list trivially satisfies global consistency due to its linear structure. Hence their approach produces a linearizable iterator for the linked list.

Petrank and Timnat were the first to develop a lock-free linearizable iterator for the linked list, though their approach cannot be naively applied to other concurrent sets. We will see that their snap-collector framework needs substantial technical changes to apply generically to sets not implemented by linked lists, so that the framework yields an iterator satisfying property (1). However, global consistency is generally not satisfied by arbitrary set implementations; for example, in a search tree, rotations may suddenly make existing elements unreachable to an active iterator. So addressing global consistency requires a completely novel approach. The main contribution of this paper is a condition called \emph{local consistency}, which a condition on set's operations that ensures global consistency. Specifically, local consistency states that an operation may never make unmodified elements unreachable to an in-progress iterator. We prove that local consistency of each set operation's atomic steps implies global consistency of any iterator's snapshot. Therefore a set implementation with locally consistent set operations can be augmented with a linearizable iterator.

We demonstrate the simplicity of our approach by verifying local consistency
for a lock-free relaxed-balance binary search by Brown \etal{} \cite{Brown2014}.
We also apply the iterator framework to two state-of-the-art set implementations:
a lock-free resizable hash table by Liu \etal{} \cite{dyn-hash}, and a lock-free unbalanced search tree by
Natarajan and Mittal \cite{fast-conc}. We evaluate the framework performance, noting an overhead of only $20-30\%$, which is comparable to the linked list iterator.

%% file: related_work.tex
Designing efficient iterators for non-blocking, linearizable concurrent data structures is a challenging research problem 
with sparse treatment in the literature. 
Petrank and Timnat \cite{Petrank2013} proposed a wait-free iterator for non-blocking
linked lists \cite{Harris2001,Timnat2012} and lock-free skip lists \cite{Herlihy2008}.
Nikolakopoulos \etal{} \cite{Nikolakopoulos2015} considered different consistency specifications 
of iterators for concurrent data structures. 
In particular, they presented both obstruction-free \cite{Herlihy2003}  and wait-free implementations of iterators 
for the classic lock-free concurrent queue by Michael and Scott \cite{Michael1996}.

Range queries and partial traversals are generalizations of 
iterations that retrieve a consistent view of one part of a data structure. 
Brown and Avni \cite{BrownA12} presented a non-blocking $k$-ary search tree with range queries. 
The range queries are only obstruction-free, as they only guarantee termination 
in the absence of concurrent updates.
Avni \etal{} \cite{Avni2013} introduced Leap-List, a concurrent data structure 
designed to provide range queries. 
Since the operations are STM-based, only obstruction freedom is guaranteed. 
Kallimanis and Kanellou \cite{KallimanisK15} proposed Dense, a wait-free concurrent graph with 
partial traversals. 
Dense provides a wait-free dynamic traversal operation that can traverse over nodes 
of the graph and return a consistent view of visited edges.

%% file: framework_arxiv.tex
We first describe the snap-collector object to establish common terminology. We then give a template defining how the set's operations are modified to support iteration. Then we outline how each operation is linearized.

\paragraph{\textup{\textbf{Snap-collector.}}} 
Due to space limitations, we discuss only the more salient aspects of Petrank and Timnat's \emph{snap-collector} object \cite{Petrank2013}. The snap-collector allows iterators to collectively build a shared snapshot of the set, and enables each set operation to notify iterators about concurrent changes that were missed. The snap-collector is set to \emph{active} when at least one iteration is taking place, otherwise it is \emph{deactivated}. 

The snap-collector maintains two linked lists: a shared \emph{snapshot-list} for iterators and individual \emph{report-lists} for set operations\footnote{We assume a bounded number of updaters in order to provide each its report-list.}. Iterators traverse the data structure according to some ordering provided by the user (depth first traversal of binary trees), and build a temporary snapshot by appending collected nodes to the snapshot-list via compare-and-swap (CAS) operations. 
Set operations add \emph{reports} to their respective report-lists via CAS, where a report contains the modified node's address and a flag indicating whether it was inserted or deleted. 
Before reporting, a set operation re-reads the snap-collector's address to check if it is active. If it is, the operation reports; otherwise it moves on without reporting anything.
Reported nodes are merged with those in the snapshot-list to produce a final snapshot: a node is in the snapshot exactly when (1) it is in the snapshot-list and an insert report, and (2) it is not in a delete report.
Multiple concurrent iterators synchronize by using the same snap-collector and thus building the same snapshot.

\paragraph{\textup{\textbf{Data structure assumptions.}}} Before applying the snap-collector to create a linearizable set iterator, we must state some basic assumptions to make the original construction generic. First, we distinguish between nodes and keys: we assume that the elements in the set, called \emph{keys}, are each contained in a level of indirection called a \emph{node}. If keys are not contained in nodes (as in a hash set), we wrap them in nodes. It is also assumed that memory allocated to nodes is not reclaimed, so that nodes are unique. These two assumption are to avoid ABA issues for keys that are inserted and deleted several times. Nodes are then endowed with a boolean \emph{mark} field that indicates logical deletion.

We assume there is predefined single-threaded procedure, called a \emph{sequential iterator}, for traversing nodes in the set. The user must provide the sequential iterator since it may be highly specific to the underlying data structure, though there is typically a natural one. For example, a binary search tree may use depth-first traversal, or a linked list may use a linear traversal. Iterators follow this procedure to traverse the set.

\vspace{-3pt}

\paragraph{\textup{\textbf{New set operations.}}} 
Here we give our template for creating new set operations that report their updates to a snap-collector. 
It generalizes that of \cite{Petrank2013}, though we make non-trivial modifications to make it general.
The set's original insert and delete operations are referred to as \dsins{} and \dsdel{}. 
The original \dsins{} is used as is, but \dsdel{} operation is altered: instead of deleting a specified key, \dsdel{} deletes a specified \emph{node}. The \seek{($k$)} operation is defined as the subroutine of the original set operations that searches for and returns the node containing key $k$, if one exists.
Together, \dsins{}, \dsdel{}, and \seek{} are used to implement new set operations compatible with a snap-collector iterator. 

\input{pseudocode.tex}

\textbf{Insert}: Start with \seek{}($k$). If $k$ not found, call \dsins{}($k$). If unsuccessful, restart. Otherwise report an insert and return \True. If node with $k$ exists, check its mark. If unmarked, report an insert return \False. Otherwise report a delete, remove node via \dsdel{}, and restart.

\textbf{Delete}: Start with \seek{}($k$). If no node $N$ containing $k$ found, return \False. Otherwise try marking $N$ with a CAS. If CAS succeeds, report a delete, remove $N$ via \dsdel{}, and return \True. If CAS fails, report a delete, call \dsdel{}($N$), and restart.

\textbf{Contains}: Start with \seek{}($k$). If no node $N$ contains $k$ found, return \False. Otherwise if $N$ is marked, report a delete and return \False. Otherwise report an insert and return \True.

\textbf{Iterate}: Read the snap-collector pointer. If not yet created, or deactivated, create/activate a new snap-collector. Traverse nodes according to sequential iteration order. Upon traversing each unmarked node, check if snap-collector is still active. If active, append node to snapshot-list via CAS. If CAS fails, retry until success or until snap-collector deactivation. After traversal finishes, deactivate snap-collector, and \emph{block} snapshot-list and report-lists by appending sentinels to them. Sort and de-duplicate nodes in snapshot-list, and merge with reports to produce final snapshot.



\subsection{Linearizability}
We present a summary of each operation's possible linearization points (LPs). The arguments are similar in spirit to Petrank and Timnat's approach \cite{Petrank2013}, though they require substantial changes to be made generic. Set operations below are either \emph{successful} (returns \texttt{True}) or \emph{unsuccessful} (returns \texttt{False}).

\textbf{Iterate.} Linearized at its snap-collector's deactivation. Recall that 
an iterator deactivates its snap-collector after it finishes iteration over the set.

\textbf{Successful insert.} Typically linearizes at the time it reports its inserted node. However, dependencies between $\ins$ and other operations can force $\ins$ to linearize at its reports. For example, if a concurrent successful $\con(k)$ finishes before $\ins(k)$ reports, then $\ins$ must linearize before $\con$ returns. Since $\con$ reports an insert, we linearize $\ins$ at the \emph{first} insert report of $k$. 

There are two cases in which $\ins(k)$ linearizes before any reporting, due to $k$ becoming visible to other operations. In the first case, a concurrent $\iter$ collects $k$ before reporting happens. So we consider the \emph{first} collect of $k$ as another possible LP. In the second case, a successful $\del(k)$ may occur between physical insertion of $k$ and its reporting. Then $\ins$ linearizes immediately before the LP of $\del$.

So a successful $\ins(k)$ linearizes at either (i) the first insert report of $k$, (ii) the first collect of $k$, or (iii) the LP of a (concurrent) successful delete of $k$ (whichever of (i), (ii), (iii) occurs first).

\textbf{Unsuccessful insert.} It reports insertion of $k$ for the same reason $\con(k)$ reports $k$, i.e., to avoid an unsuccessful insert linearizing before a successful insert. There are two possible LPs. If there is a concurrent successful $\ins(k)$, the unsuccessful $\ins(k)$ linearizes immediately after the LP of successful $\ins(k)$. Otherwise, $k$ exists, and there is no concurrent successful $\ins(k)$, but a $\del(k)$ trying to delete it. We linearize insert when it is sure that $\del(k)$ has not already removed the node $N$ containing $k$, which is when $\ins$ sees that $N$ is unmarked. 

\textbf{Successful delete.} Typically linearizes at the \emph{first} delete report of $k$. However, this may not always suffice: a concurrent iterator can see the node marked, ignore it, and finish before any delete report is generated. Therefore, if \emph{all} concurrent iterators see $N$ with key $k$ has been marked, then delete is linearized at the time it marks $N$. Otherwise, it is linearized at the first delete report.

\textbf{Unsuccessful delete.} Typically linearizes when its $\seek(k)$ indicates that $k$ was not found. However, a concurrent successful $\ins(k)$ might return before $\seek$ returns, forcing $\del(k)$ to linearize before $\ins$ returns. So the LP of unsuccessful $\del(k)$ is (i) immediately before the LP of concurrent successful $\ins(k)$, or (ii) when $k$ is not found (whichever occurs first).

\textbf{Successful contains.} A successful $\con(k)$ behaves similarly to unsuccessful $\ins(k)$, so the possible LPs of $\con(k)$ are the same as those of an unsuccessful $\ins(k)$.

\textbf{Unsuccessful contains.} There are two cases. Either the operations is unable to find $k$, or it sees that node $N$ containing $k$ is marked, or logically absent. In the first case, the LPs are the same as an unsuccessful $\del(k)$. In the second case, if $\con(k)$ overlaps with the LP of a successful $\del(k)$, then $\con(k)$ linearizes immediately after. Otherwise, $\con(k)$ is linearized at the start of its execution. The LP of $\del(k)$ must be before the $\con$ operation, because $\con(k)$ sees that $N$ is marked.

\vspace{10pt}

 The argument above is enough to ensure that iteration is consistent with concurrent set operations. However in general, this approach overlooks the possibility of an iterator failing to
 collect nodes not modified during its execution. That is, a linearizable iterator must also be \emph{globally consistent}. The linked list iterator trivially satisfies global consistency, though this property does not necessarily hold for more complex data structures. Moreover, global consistency may be difficult to verify, so in the next section we introduce a property that is easier to work with, called local consistency.

%

%% file: pseudocode.tex
\begin{algorithm}[H]  
\caption{Framework operations}
\label{fig:alg-1}
\begin{multicols}{2}
\begin{algorithmic}[1]


\Procedure{$\ins$}{$k$}
	\State $n \larr \seek(k)$;
	\If {$n$ exists, i.e., ($k$ is found)} 
		\If {$n$ is marked} 
			\State Read snap-collector sc;
			\State Report deletion of $n$;
			\State $\dsdel(n)$;
			\State $\ins(k)$;
		\Else 
			\State $\tryReport(n)$;
			\State return $\False$;
		\EndIf
	\Else 
		\State $b \larr \dsins(k)$;
		\If {$b = 1$} 
			\State $\tryReport(n)$;
			\State return $\True$;
		\Else 
			\State $\ins(k)$;
		\EndIf
	\EndIf
\EndProcedure

\Statex

\Procedure{$\tryReport$}{$n$}
	\State Read snap-collector sc;
	\If {$n$ is marked} 
		\State Report deletion of $n$;
	\Else 
		\State Report insertion of $n$;
	\EndIf
\EndProcedure

\Statex

\newpage

%

\Procedure{$\del$}{$k$}
	\State $n \larr \seek(k)$;
	\If {$n$ exists} 
		\State $b \larr \mark(n)$;
		\If {$b = 0$} 
			\State Read snap-collector sc;
			\State Report deletion of $n$;
			\State $\dsdel(n)$;
			\State $\del(k)$;
		\Else 
			\State Read snap-collector sc;
			\State Report deletion of $n$;
			\State $\dsdel(n)$;
			\State return $\True$;
		\EndIf
	\Else 
		\State return $\False$;
	\EndIf
\EndProcedure
\setcounter{algorithm}{1}

\Statex

\Procedure{$\con$}{$k$}
	\State $n \larr \seek(k)$;
	\If {$n$ exists} 
		\If {$n$ is marked} 
			\State Read snap-collector sc;
			\State Report deletion of $n$;
			\State return $\False$
		\Else 
			\State $\tryReport(n)$;
			\State return $\True$;
		\EndIf
	\Else 
		\State return $\False$;
	\EndIf
\EndProcedure


\end{algorithmic}
\end{multicols}
\end{algorithm}

%% file: theory_arxiv.tex

\section{Local and global consistency}
\label{section:locality}

We describe \emph{local consistency} and its role in supporting linearizable set iterators. As before, we are given a sequential iterator for a set. One would hope that the sequential iterator functions correctly in multithreaded settings, though this often fails to be true. For instance, the sequential iterator that traverses the internal self-balancing tree in Figure \ref{fig:rotation} may see an inconsistent snapshot. In particular, suppose it runs a depth-first traversal of the tree, starting at the root node (node $6$ in Figure \ref{fig:rotation}). The iterator then pauses, and another thread performs a rotation. If the iterator then continues in isolation, it fails to collect nodes $L = \{1, 2, 3, 4\}$ in the left half of the tree, since it believes $6$ is still the root. There is no way to linearize this iterator's execution since $L$ must be included in its snapshot. It is therefore essential for a linearizable iterator to collect unmodified nodes in the set, where \emph{modified nodes} are those that are either inserted or deleted during the iterator's execution. We state this property as a definition.

\begin{definition}[global consistency]
Consider an iterator over a set. Then the iterator's snapshot is \emph{globally consistent} if it contains all nodes in the set not modified by concurrent operations.
\end{definition}

Global consistency is easy to verify for the linked list iterator, though it is a more difficult property to analyze for more complex data structures. So we also introduce a property called \emph{local consistency}, a constraint on the \emph{atomic} steps of set operations, which informally states that atomic steps never make unvisited nodes unreachable to an iterator. Local consistency does not restrict basic data structure geometry, though it does constrain how set operations can move nodes around within the data structure. Verifying local consistency is more straightforward since it only requires one to work with the set operation's atomic steps. But more importantly, we show that local consistency of each atomic step implies global consistency of the iterator's snapshot.

\subsection{Views and local consistency}

Consider a fixed set implementation
(binary search tree, hash table, linked list), and let $\Oh$ denote the collection of all such data structures.
Each data structure $T \in \Oh$ is comprised of nodes storing keys. When speaking in general terms, ``node" refers to only those that contain actual keys, as opposed to e.g. internal nodes that are only used for traversal. Let $<_T$ denote the order on the node set of $T$ in which nodes are visited by a sequential iterator. This does not necessarily coincide with the set's ordering; e.g. a hash set may not even have an ordering. 

\begin{figure}[t]

\begin{subfigure}{0.12\textwidth}

\raggedleft
\includegraphics[scale=0.17]{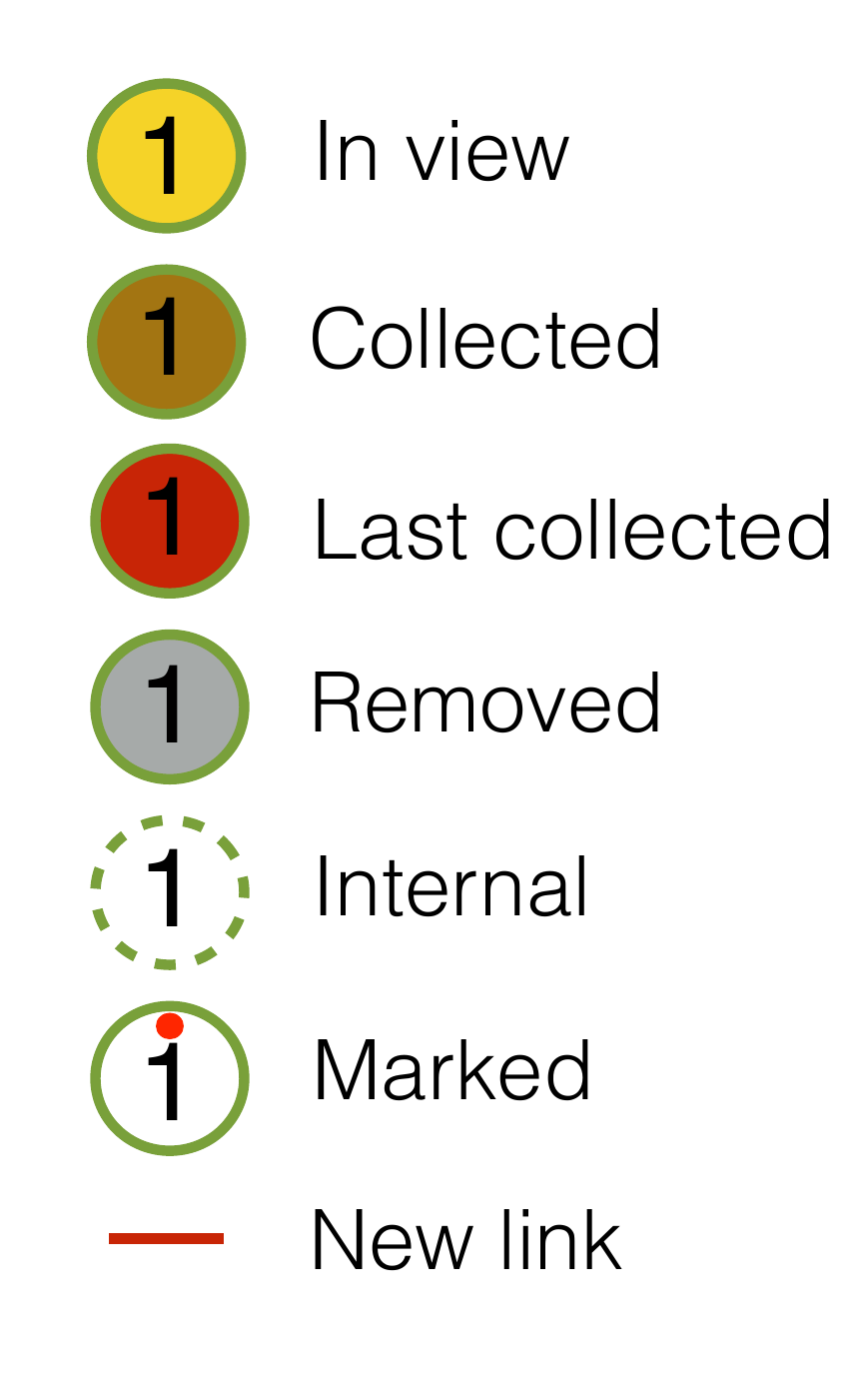}
    \label{fig:legend}
\end{subfigure}
\begin{subfigure}{0.32\textwidth}
\centering
\includegraphics[scale=0.2]{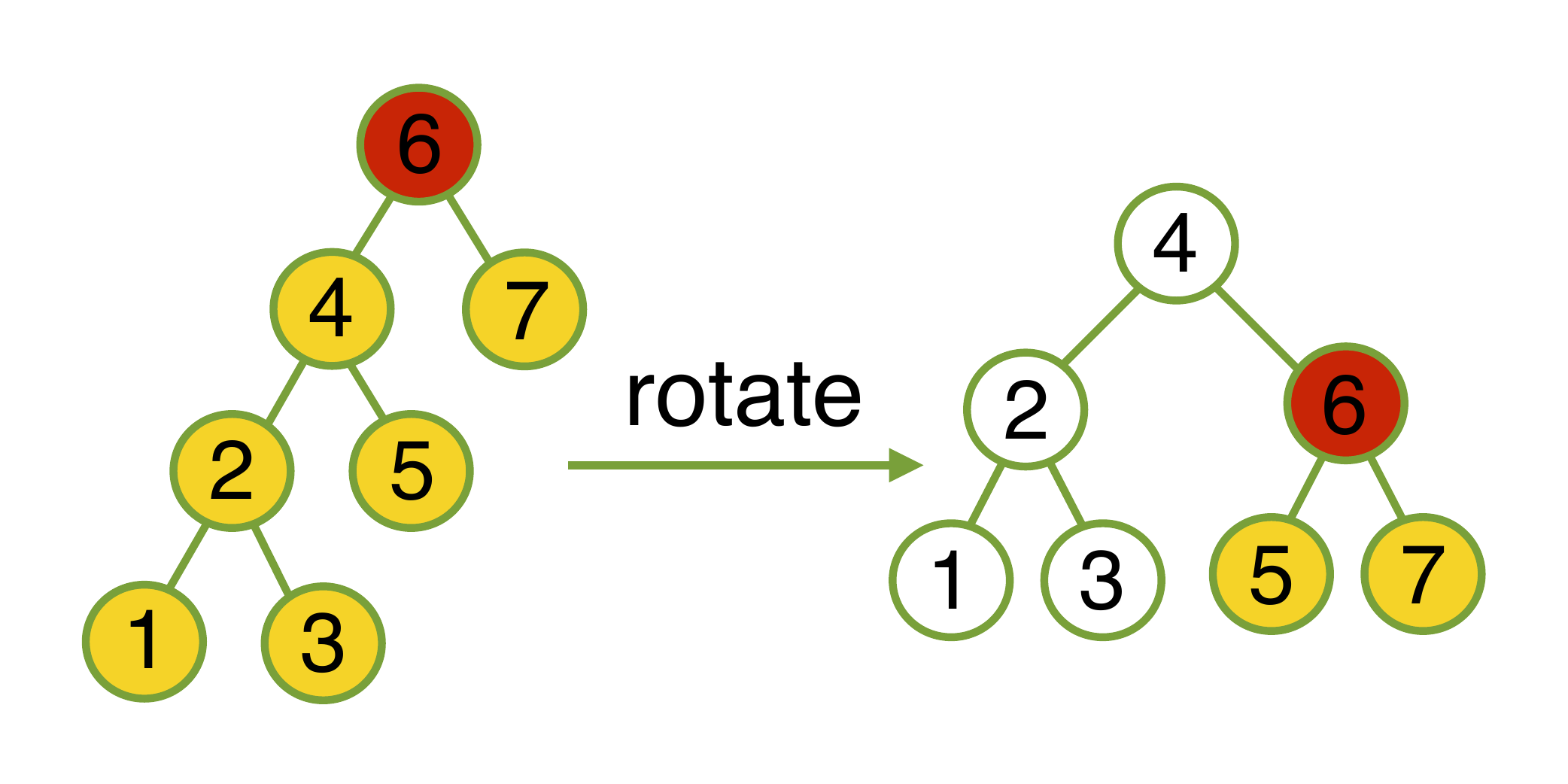}
\caption{Example rotation that is not locally consistent.}
    \label{fig:rotation}
\end{subfigure}
\begin{subfigure}{0.4\textwidth}
\centering
\includegraphics[scale=0.15]{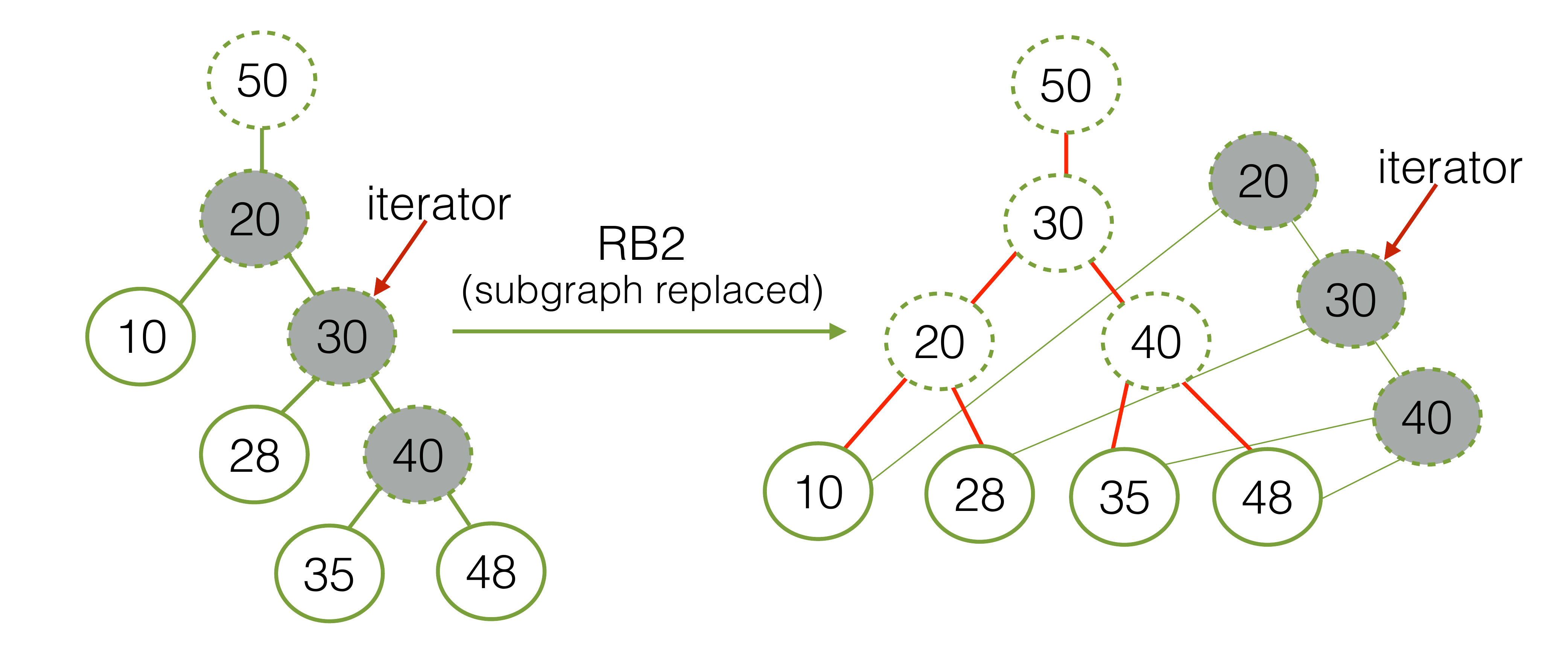}
\caption{A rotation in a chromatic tree.}
    \label{fig:RB2}
\end{subfigure}

\begin{subfigure}{.27\textwidth}
  \centering
  \includegraphics[scale=0.15]{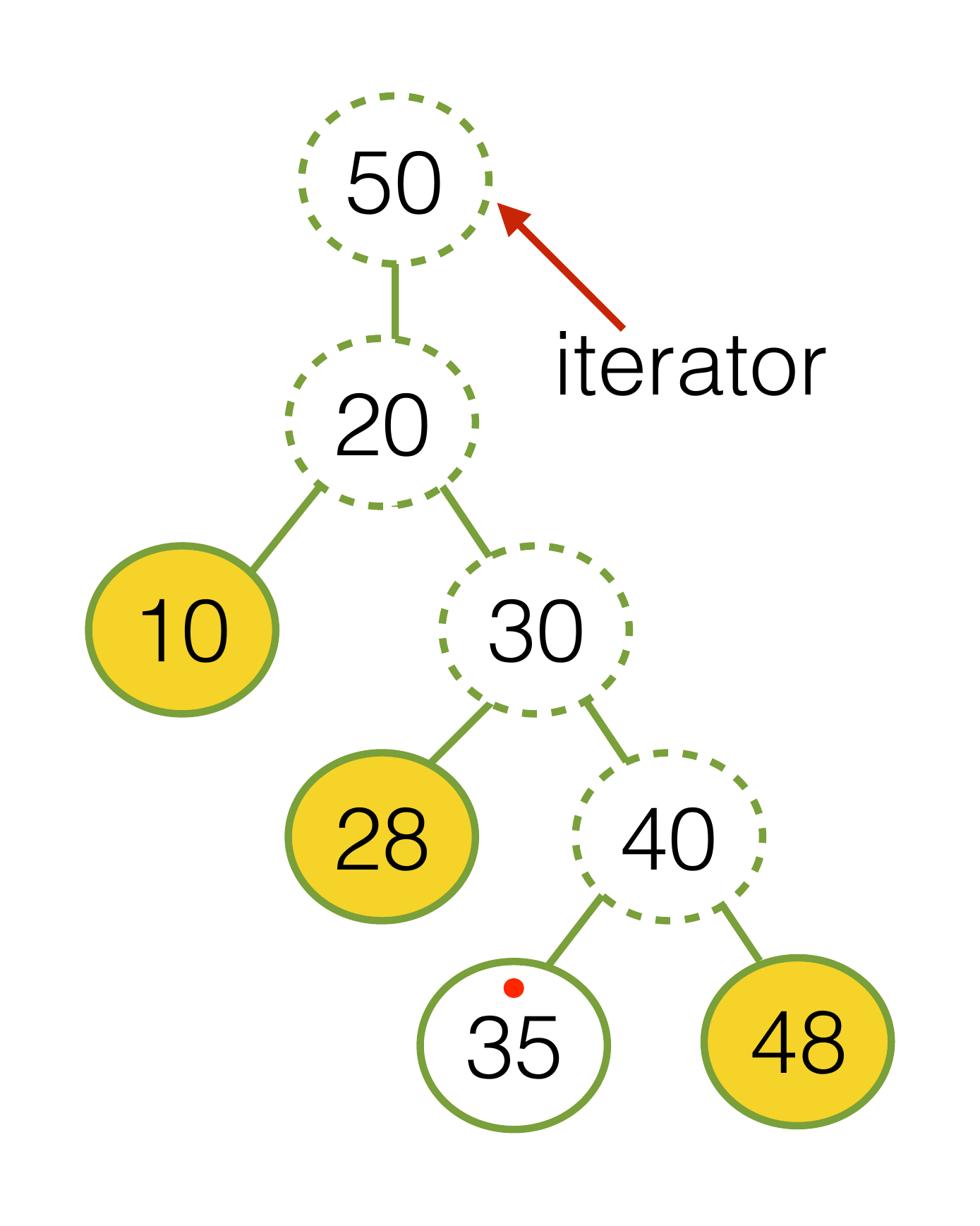}
  \caption{$V(N_0, T)$, the view of an iterator beginning its traversal.}
  \label{fig:view1}
\end{subfigure} \hfill
\begin{subfigure}{.27\textwidth}
  \centering
  \includegraphics[scale=0.15]{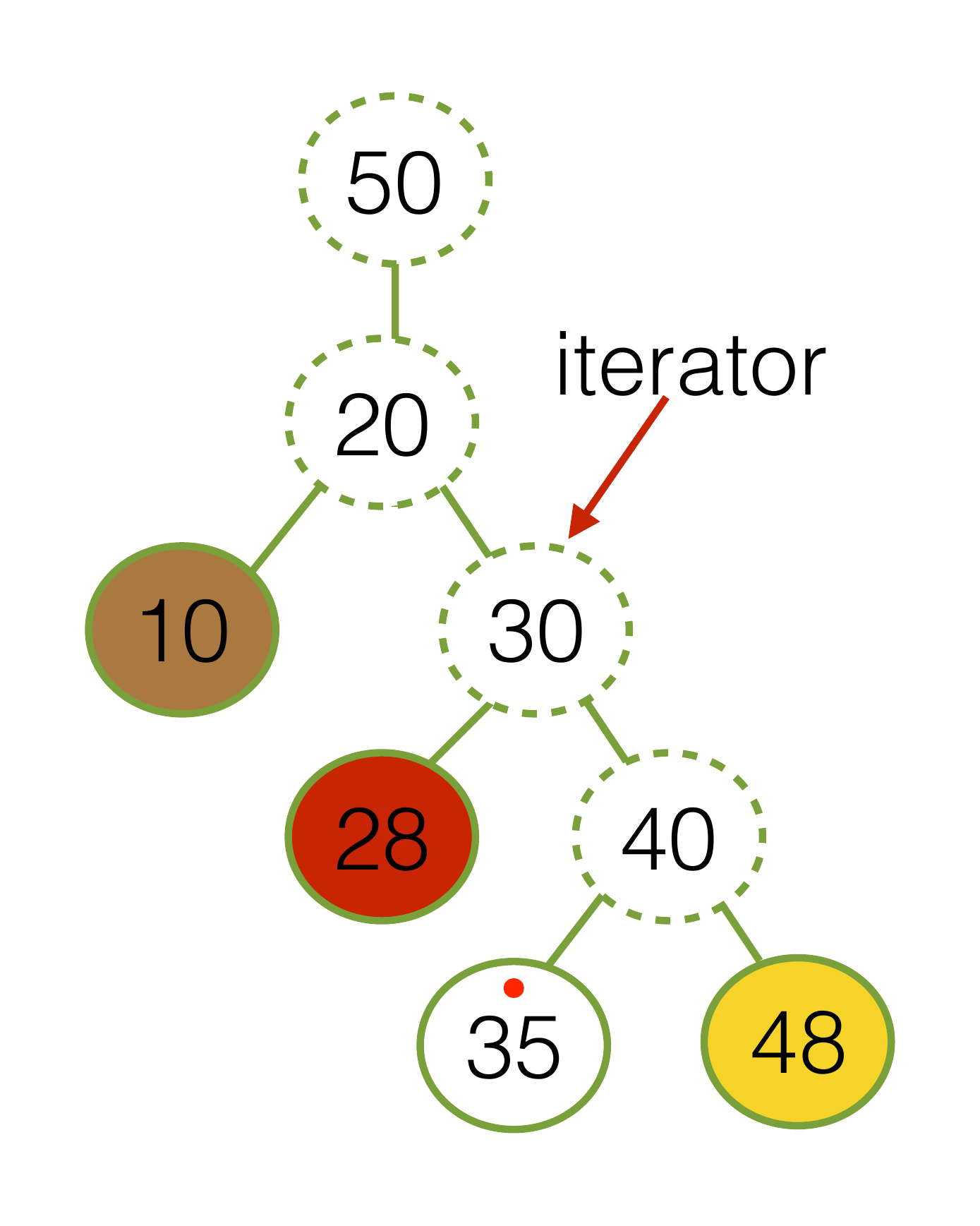}
  \caption{$V(28, T)$, the iterator's view after collecting nodes 10 and 28.}
  \label{fig:view2}
\end{subfigure} \hfill
\begin{subfigure}{.41\textwidth}
   \centering
   \includegraphics[scale=0.15]{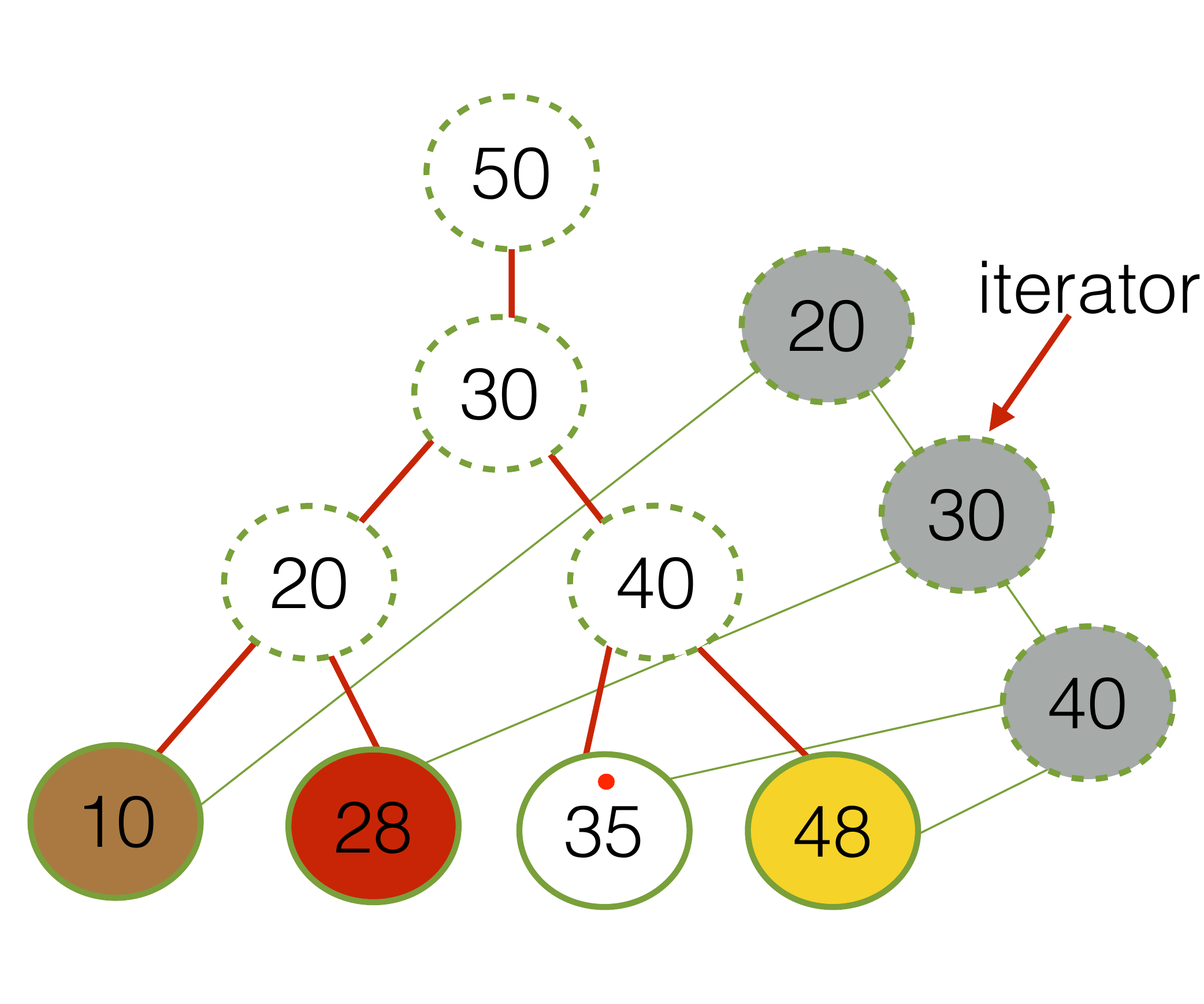}
   \caption{A rotation happens, resulting in tree $T'$. The iterator's view $V(28, T')$ remains unchanged.}
\end{subfigure}
\caption{Examples of views.}
    \label{fig:views}

\end{figure}

As a running example for this section, we use the chromatic tree of Brown \etal{} \cite{Brown2014}, which is a relaxed-balance variant of a red-black tree that implements non-blocking linearizable set operations. At a high level, each update to the chromatic tree is made in a copy-on-write manner: a connected subgraph is atomically replaced with a newly-created connected subgraph reflecting the update to the tree. Keys are only stored in the leaf nodes; data contained in internal nodes are for traversal. Note it is assumed that internal nodes in a removed subgraph persist in memory long enough for concurrent iterators located in the subgraph to finish their traversal.\footnote{This holds true in garbage collected languages.} Figure \ref{fig:RB2} illustrates a rebalancing step (which Brown \etal{} call RB2), after which a concurrent iterator may finish its traversal.

Consider an iterator traversing through data structure $T$. Then its \emph{view} is defined as follows.

\begin{definition}[view]
\label{def:view}
Let $N$ be a node in $T$. Then the \emph{view} of $T$ relative to $N$ is
defined to be the set of nodes $V(N, T) = \{M : M >_T N, \; M \textrm{ not marked} \}$. 
\end{definition}

That is, if node $N$ is most recently visited by the iterator, then $V(N, T)$ is the remaining set of nodes that the iterator would traverse were it to run solo. We take $V(N_0, T)$ as notation for the set of all nodes in $T$. See Figure \ref{fig:views} for examples of an iterator's view in a chromatic tree.

We reason about set operations by decomposing them into their atomic steps which we also call \emph{mutators}; these terms are used interchangeably. 

\begin{definition}[mutator]
\label{def:mutator}
A \emph{mutator} is an atomic step, modeled as a function $m: \Oh \to \Oh$ that
maps each data structure $T$ to a modified data structure $m(T)$.

\end{definition}

Marking, physical insertion, and physical removal of a node via CAS are examples of mutators. 
The RB2 operation in Figure \ref{fig:RB2} is an atomic step which we denote by mutator $m_{RB2}$. We now define local consistency, which restricts how mutators can change an iterator's view.

\begin{definition}[local consistency]
\label{def:locality}
Let $m$ model an atomic step in a set operation \textbf{op}, and suppose \textbf{op} modifies set of nodes $R$. Then $m$ is called \emph{locally consistent} if for any iterator, any data structure $T \in \Oh$ and any $N \in T$, the following holds (where $\triangle$ is the symmetric set difference):
\[ V(N, T) \; \triangle \; V(N, m(T)) \subseteq R \]
\end{definition}

An operation composed from locally consistent atomic steps is also called \emph{locally consistent}. In the above definition, $R$ is called the \emph{change set} of $m$, and may be written as $R^m$ to denote its association with $m$. If $m$ is an atomic step of inserting node $N$, then $R^m = \{N\}$, or if it is a step in deleting $M$, then $R^m = \{M\}$. The $m_{RB2}$ mutator never changes an iterator's view, as in Figure \ref{fig:RB2}, so $V(N, T) \; \triangle \; V(N, m_{RB2}(T))$ is always empty. So $m_{RB2}$ is locally consistent.

\begin{figure}[t]
\centering
\begin{subfigure}{.48\textwidth}
\centering
\includegraphics[scale=0.15]{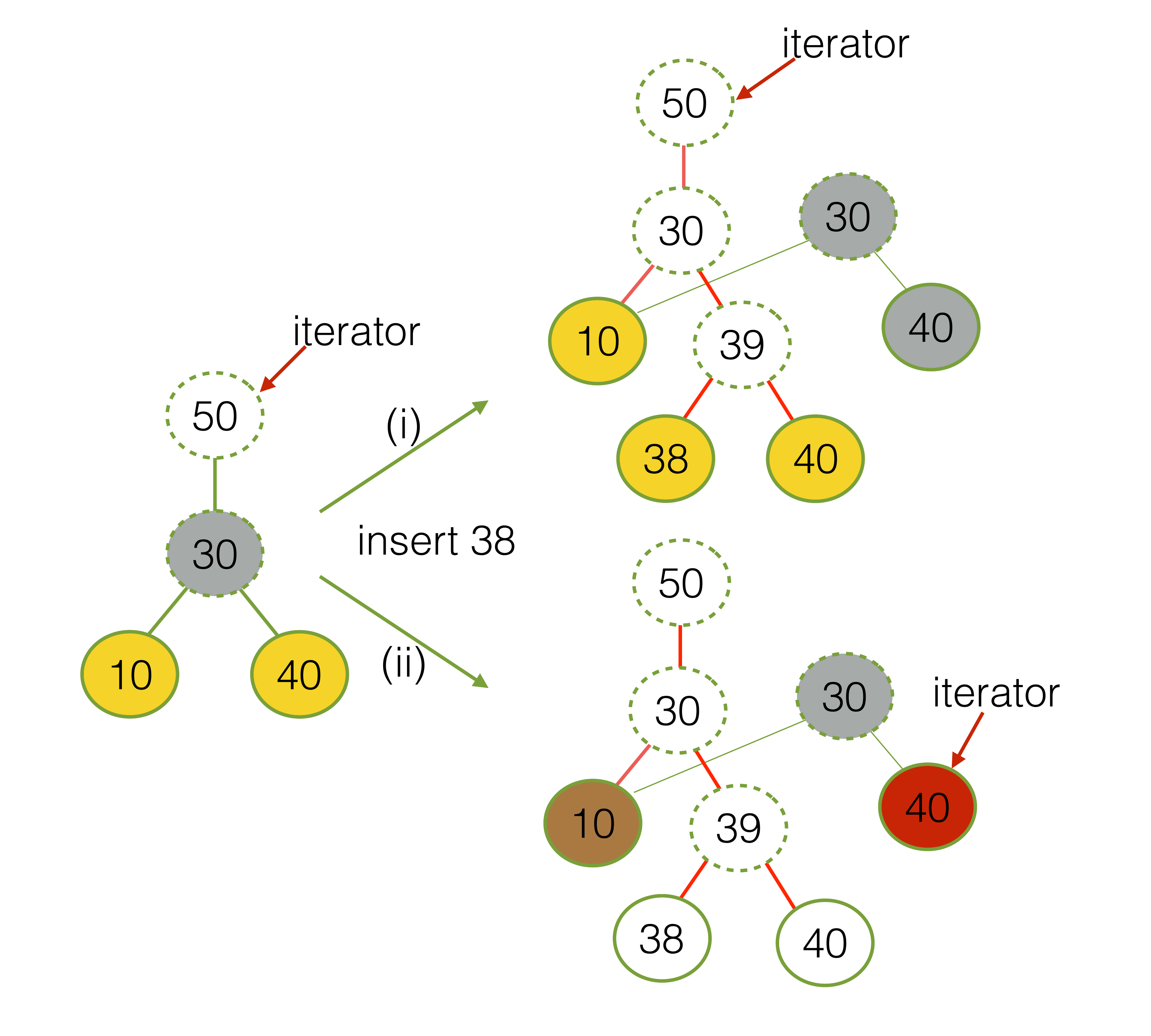}
\caption{Executions (i) and (ii) of \ins{(38)}.}
\label{fig:insert}
\end{subfigure}\hfill
\begin{subfigure}{.48\textwidth}
\centering
\includegraphics[scale=0.15]{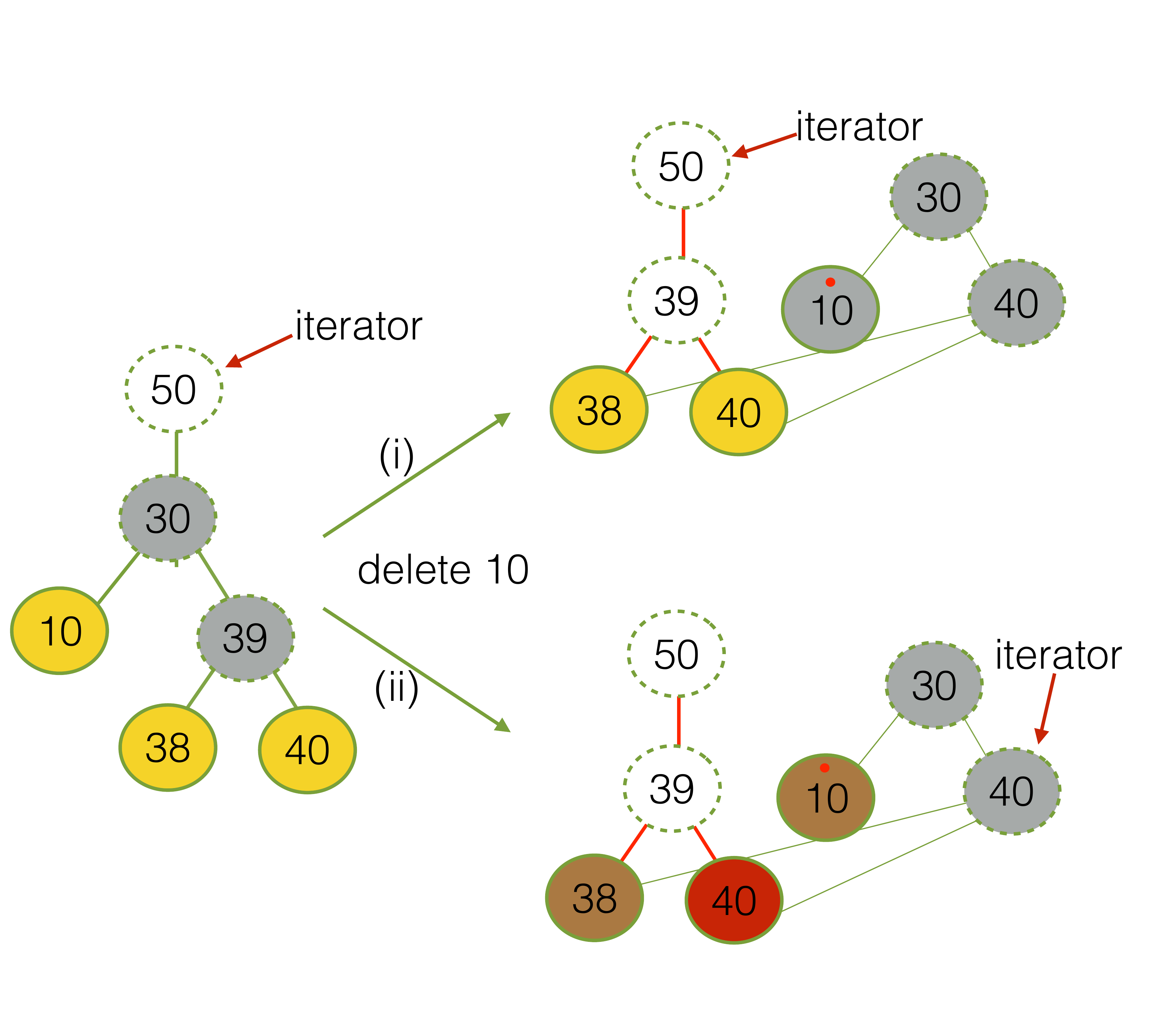}
\caption{Executions (i) and (ii) of \del{(10)}.}
\label{fig:delete}
\end{subfigure}
\caption{\ins{} and \del{} operations, each in the presence of an iterator.}
\label{fig:updates}
\end{figure}

Figure \ref{fig:updates} illustrates the atomic steps of insert and delete operations that modify the chromatic tree. Let $m_i$ denote the mutator representing \ins{}'s atomic step. In Figure \ref{fig:insert}, an \ins{(38)} runs concurrently with an iterator. The iterator will either collect or miss node 38, depending on whether the node is added before the iterator reaches it. Local consistency allows for the uncertainty of 38 being in the iterator's view, but requires nodes 10 and 40 to remain in the view if they were not previously collected. This is true for each interleaving of an iterator with \ins{}, so $V(N, T) \triangle V(N, m_i(T)) \subseteq \{38\}$. So $m_i$ is local.

In Figure \ref{fig:delete}, \del{(10)} runs concurrently with an iterator. There are two atomic steps, marking and physical removal of 10, which both have a change set of $\{10\}$. The marking mutator $m_\mu$ removes 10 from the iterator's view, since marked nodes are skipped by iterators. The iterator will either collect or miss 10, depending on whether it is marked before being collected. As with insertion, local consistency permits this uncertainty, so $m_\mu$ is locally consistent. The removal mutator $m_d$ does not logically delete any nodes since 10 is already marked, so similar reasoning to RB2 shows that $m_d$ never changes an iterator's view. So $m_d$ is also locally consistent.

For each mutator $m$, the change set $R^m$ is added to an aggregate set $R$ over the span of an iterator's execution. As $R$ grows with more concurrent mutators, the iterator may miss more nodes. However, local consistency of each $m$ implies \emph{global consistency}, or that iterators never miss unmodified nodes in the set. This is stated as a theorem.

\begin{theorem}\label{thm:local}
Let $T$ be a set data structure with a given sequential iterator, and locally consistent set operations. Then any iteration over $T$, possibly concurrent with other operations, produces a globally consistent snapshot.
\end{theorem}

We first provide some intuition, followed by a proof. Consider a data structure $T$ whose set operations' atomic steps are locally consistent. Take any execution of an iterator over $T$, possibly concurrent with other operations, and decompose the entire execution into atomic steps. Then at each step, local consistency guarantees that no unmodified node ever leaves the iterator's view, unless it has already been collected by the iterator. Inducting over the execution's atomic steps shows that the same holds true at the end of the iterator's execution, which implies that the iterator must have collected every unmodified node, or global consistency.

To prove the converse, suppose one of the set operation's atomic steps $m$ is not local. Then there is an execution where $m$ shifts an unmodified node $N$ outside of a concurrent iterator's view. Extending this execution by letting the iterator run solo, the iterator returns a snapshot that is globally inconsistent since it does not contain $N$.

\begin{proof}

Let $m_i$ be locally consistent mutators with induced change sets $R_i$, running concurrently with an iterator.
The iterator's execution occurs over a sequence of time steps $t_1, \ldots t_n$, as specified by the mutators $m_1, \ldots m_n$. Let $t_0$ denote the starting time of the iterator's execution, and define $T_0 = T$. Then we inductively define $T_{i+1} = m_{i+1}(T_i)$, so that
$T_{i}$ is the data structure after the first $i$ mutators are applied to $T$. Let $U = V(N_0, T)$, the set of all nodes initially contained in the data structure.

Since each mutator represents an atomic operation, the iterator runs in isolation between each $t_i$ and $t_{i+1}$. Let $N_i$ denote the node most recently visited in $T_i$ up to time $t_i$, and let $S_i$ be the set of all nodes collected by the iterator up to time $t_i$. Then the following holds:
\[ S_{i + 1} = S_i \cup (V(N_i, T_i) \setminus V(N_{i+1}, T_i) ) \]
This is because the set of nodes that the iterator collects between times $t_i$ and $t_{i+1}$ is precisely $V(N_i, T_i) \setminus V(N_{i+1}, T_i)$.

Let $X_i$ denote the set of unmodified nodes up to time $t_i$. As time progresses, an increasing number of nodes in the data structure are modified by mutators, so $X_i$ shrinks over time. Defining $\ol{R}_i = \bigcup_{j \le i} R_j$ to be the aggregate of all modified nodes up to time $t_i$, we see that $X_i = U \setminus \ol{R}_i$. The theorem we wish to prove may be stated as follows:
\[ X_n \subseteq S_n \]
Since $X_n$ is the set of nodes not modified by a mutator at the end of the iterator's execution, each node in $X_n$ should also be in $S_n$, which is the set of all nodes the iterator collects over its execution. We prove this by inducting on $t_i$. For any $t_i$, we want to prove the following:
\[ X_i \subseteq S_i \cup V(N_i, T_i) \]
That is, at any given time, any unmodified node has already been collected by the iterator, or will be collected at a future time.
To begin the proof, we start with the base case $i = 0$, which states that $X_0 \subseteq S_0 \cup V(N_0, T_0)$. But $S_0 = \varnothing$ since no nodes have been collected at the start of the execution, and $V(N_0, T_0) = U$. Also, by definition, $X_0 = U \setminus R_0$. The containment $U \setminus R_0 \subseteq U$ clearly holds, so the base case $X_0 \subseteq S_0 \cup V(N_0, T_0)$ is true.

Next, inductively assuming that $X_i \subseteq S_i \cup V(N_i, T_i)$ holds, we want to show that the same is true for $i+1$. For brevity, we write $Y_i = S_i \cup V(N_i, T_i)$, so that $X_i \subseteq Y_i$. The set $Y_i$ contains nodes the iterator collects if it runs concurrent with mutators up to time $t_i$, then runs in isolation from $t_i$ onwards. We know that $X_i \setminus X_{i+1} = R_{i+1}$, so if we show that $Y_i \setminus Y_{i+1} \subseteq R_{i+1}$, then we can conclude that $X_{i+1} \subseteq Y_{i+1}$. Intuitively, we want to show that more nodes are removed from $X_i$ (to obtain $X_{i+1}$) than are removed from $Y_i$ (to obtain $Y_{i+1}$). Calculating $Y_i \setminus Y_{i+1}$, we have:
\[ Y_i \setminus Y_{i+1} = (S_{i} \cup V(N_{i}, T_{i})) \setminus (S_{i+1} \cup V(N_{i+1}, T_{i+1})) \]
\[\subseteq (S_{i} \cup V(N_{i}, T_{i})) \setminus (S_{i} \cup V(N_i, T_i) \setminus V(N_{i+1}, T_i) \cup V(N_{i+1}, T_{i+1})) \]
\[\subseteq V(N_{i+1}, T_{i}) \setminus V(N_{i+1}, T_{i+1}) \subseteq V(N_{i+1}, T_{i}) \triangle V(N_{i+1}, m_{i+1}(T_i)) \]
We get $Y_i \setminus Y_{i+1} \subseteq V(N_{i+1}, T_{i}) \triangle V(N_{i+1}, m_i(T_i))$. Finally, invoking local consistency of $m_{i+1}$, we arrive at $Y_i \setminus Y_{i+1} \subseteq R_{i+1}$. Therefore $X_{i+1} \subseteq Y_{i+1} = V(N_{i+1}, T_{i+1})$. So by induction, we have $X_n \subseteq S_n \cup V(N_n, T_n)$, and since $V(N_n, T_n) = \varnothing$, we have $X_n \subseteq S_n$. This completes the proof of the theorem.

\end{proof}

It follows that local consistency at each atomic step implies global consistency of the entire iteration. Here we have proved locally consistency of a few atomic steps of the chromatic tree; the remaining steps are shown to be locally consistent using similar arguments. The other atomic steps, along with local consistency proofs for the two other sets (unbalanced binary search trees and hash sets) can be found in the next section. Then we evaluate the framework applied to the unbalanced binary search tree and the hash set.

%% file: applications_appendix.tex
\section{Applying the Framework}
We apply our framework to data structures taken from the literature, and prove their set operations to be locally consistent

\subsection{Lock-Free Binary Search Tree}
We consider the lock-free binary search tree of Natarajan and Mittal \cite{fast-conc}. In this tree, keys
are stored in leaf nodes, and each internal node maintains \textit{left} and \textit{right} fields of an internal node, 
which are \emph{edges} pointing to the respective child nodes. Each edge can be marked as \emph{flagged} or \emph{tagged}. 
A flagged edge signifies that both the nodes incident to the edge are to be removed from the tree, 
and a tagged edge indicates that only the parent node will be removed. Such nodes are considered logically deleted.
Once an edge is flagged or tagged, it is made immutable, so it cannot point to another node. 

\subsubsection{Operations}

\begin{figure}
\centering
\includegraphics[width=.6\linewidth]{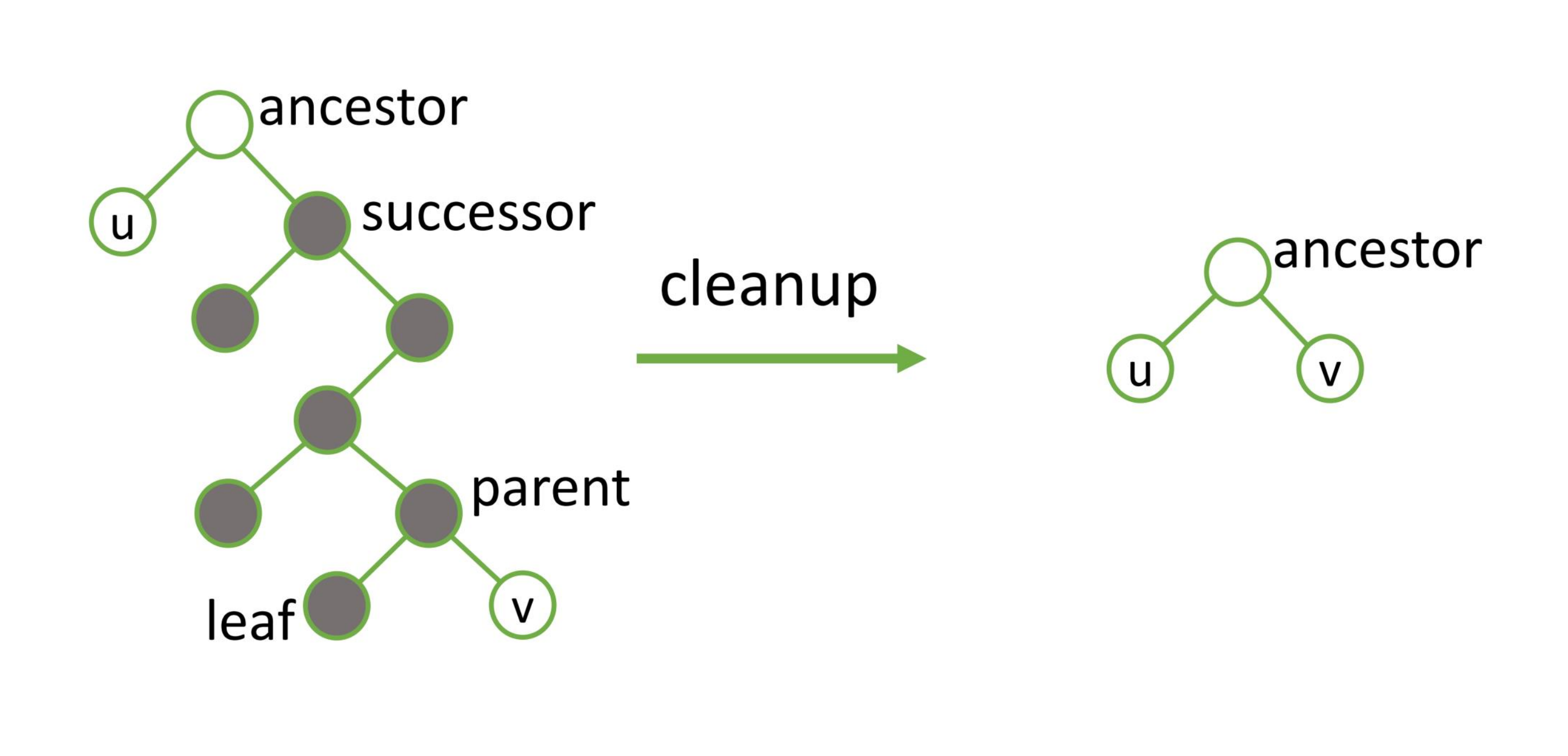}
\caption{A single CAS that physically removes all shaded nodes.}
\label{fig:unbalanced}
\end{figure}

Each set operation starts with a read-only \dsseek($key$) subroutine 
\footnote{To resolve terminology conflicts, we use \dsseek{} to refer to the \seek{} function in the original algorithm.}
that uses a binary search to attempt findind a leaf with \textit{key} (see Algorithm 1 in \cite{fast-conc}). The search terminates
at a leaf node, regardless of whether that leaf contains the correct key. \dsseek{} returns a \textit{seekRecord}
which consists of this \textit{leaf} node, the \textit{parent} of the leaf, 
the parent's nearest ancestor that is not tagged, 
called the \textit{ancestor}, and the ancestor's child in the path, called the \textit{successor}. 
See Figure \ref{fig:unbalanced} for an example of nodes in a \textit{seekRecord}.

Set operations sometimes call a \clean{} subroutine to help physically remove nodes 
that have been logically but not physically deleted (see Algorithm 4 in \cite{fast-conc}). 
The \clean{} function takes as arguments a \textit{key} and 
the \textit{seekRecord} returned by \dsseek(\textit{key}) called by operation earlier. 
As shown in Figure \ref{fig:unbalanced}, 
\clean{} attempts to physically remove the \textit{leaf} node in \textit{seekRecord}
(or the sibling of \textit{leaf}, depending on which of them is marked) 
and all nodes in the path between \textit{successor} and \textit{parent}. This is accomplished by applying a CAS to
swing an edge (\textit{left} or \textit{right}) of \textit{ancestor} to point from \textit{successor} to \textit{leaf}'s sibling (or \textit{leaf}),  
since each nodes along the path are tagged and hence logically deleted. \clean{} returns true if the CAS succeeds and false otherwise.

The \con(\textit{key}) operation calls \dsseek{($key$)}, retrieves \textit{leaf} from the returned \textit{seekRecord}, and compares the key in \textit{leaf} to $key$.If they match, the operation returns true, otherwise it returns false.

The \ins(\textit{key}) operation begins with \dsseek{($key$)}, and returns false immediately if \textit{key} matches the key of \textit{leaf} returned by the \dsseek{}, as a node of the same key is already present in the tree.  
Otherwise, as shown in Figure \ref{fig:unbalanced}, 
it makes a single CAS to attempt to insert a new internal node as a child of \textit{parent}, 
with a new leaf node of \textit{key} and \textit{leaf} as its two children (by linking them before the CAS).  
If the CAS fails, there must be a conflicting \del{} operation there, and therefore the \ins{} operation will 
call the \clean{} function if necessary to help physically remove nodes and then retry from the beginning. 

The \del(\textit{key}) operation begins with \dsseek{($key$)}, (Algorithm 3 in \cite{fast-conc}) and returns false 
if its key does not match the key of \textit{leaf} returned by its \dsseek(\textit{key}) subroutine. 
This is because a node of the same key is not present in the tree. 
Otherwise, \del{} tries to logically delete \textit{leaf} by making a CAS to flag the edge between \textit{parent} and \textit{leaf}. 
If the CAS succeeds, the \del{} operation will call \clean{} repeatedly
until \textit{leaf} has been physically removed (lines 83-87 of Algorithm 3 in \cite{fast-conc}). 
(Before each \clean{} call, a \dsseek(\textit{key}) is also called to get an updated \textit{seekRecord}).
If the CAS fails to flag the edge because of a conflicting update, 
\del{} will first call \clean{} if necessary to help physically remove other marked nodes,
and then restart its own \del{} operation from the beginning. 

\subsubsection{Local consistency}

We prove local consistency of each operation's atomic steps. Recall that leaf nodes are the only
ones that store keys, so iterators' views are only taken with respect to leaf nodes. 
There are only two operations that may change the view of an iterator. 
One is the CAS operation in \ins{}(\textit{u}) that swings a child pointer of a \textit{parent} 
to a new internal node \textit{v}, where \textit{v} has as children \textit{u} and \textit{parent}'s original child \textit{leaf}.
The change set of this CAS is $R^m = \{u\}$, since $\{u\}$ is the node being inserted.
Leaves in the tree other than \textit{leaf} and \textit{u} are not affected by the CAS,
so we only have to ensure that \textit{leaf} remains in an iterator's view, given that $u \in R^m$.
If an iterator has not visited \textit{parent} yet, it will be able to reach \textit{leaf} 
either following the old path $\textit{parent} \rightarrow \textit{leaf}$ before the CAS, 
or the new path $\textit{parent} \rightarrow v \rightarrow u$ after the CAS. 
If the iterator is currently at \textit{parent},  it will be able to reach \textit{leaf}
via the path $\textit{parent} \rightarrow \textit{leaf}$ both before and after the CAS, 
as it has already read \textit{parent} with the old edge (\textit{parent}, \textit{leaf}). 
In either case, \textit{leaf} remains in the iterator's view.
Therefore, we have that for each (leaf) node $N$, and any iterator, $V(N, T)\Delta V(N, m(T)) \subseteq \{u\} = R^m$.
So insert's CAS is locally consistent.

The other atomic step is the CAS in \clean{} that swings a child pointer of a node \textit{ancestor} to a leaf \textit{v}
in order to remove a connected group of nodes between them, as in Figure \ref{fig:unbalanced}.
As stated earlier, all nodes removed by the CAS were already marked as logically deleted by previous \del{} operations,
so those nodes do not affect the view of an iterator. Now we consider node \textit{v},
the only unmarked leaf whose reachability might be affected by the CAS. 
If an iterator is at any node in the subtree rooted at \textit{successor}, 
$v$ is reachable following the path from \textit{ancestor} to \textit{parent} in the old tree, 
both before and after the CAS. 
If an iterator has not visited $successor$ yet, it can reach $v$ following either the same path 
before the CAS, or the path $ancestor \rightarrow v$ directly after the CAS.
In either case, $v$ remains in the iterator's view. 
Therefore $V(N, T) \Delta V(N, m(T)) = \varnothing$ and hence this CAS is also locally consistent.

\subsection{Lock-Free hash set}
\label{sec:hash_set}
Liu \etal{} \cite{dyn-hash} designed a lock-free, dynamically resizable hash set. The hash set is implemented as a linked list of \hnode{} objects, each representing a version of the hash set. Each \hnode{} contains an array of buckets implemented from \fset{}s, which are sets that can be \emph{frozen}, or made permanently immutable. Each \hnode{} also has a $pred$ pointer to the previous \hnode{}. The $head$ \hnode{} is the current version.

Set containment is performed by checking the appropriate bucket for the queried key. A thread inserts or deletes a key in a bucket by creating a copy of the bucket, updating the copy with the new key, and replacing the original bucket with a CAS. The hash set's operations periodically trigger a resize, in which an \hnode{} is created with new array of buckets, and buckets are initialized and copied over lazily. The $head$ pointer is swung to the new $\hnode{}$, and old buckets are frozen when copies are created for the new \hnode{}.

\begin{figure}
\centering
\includegraphics[width=.7\linewidth]{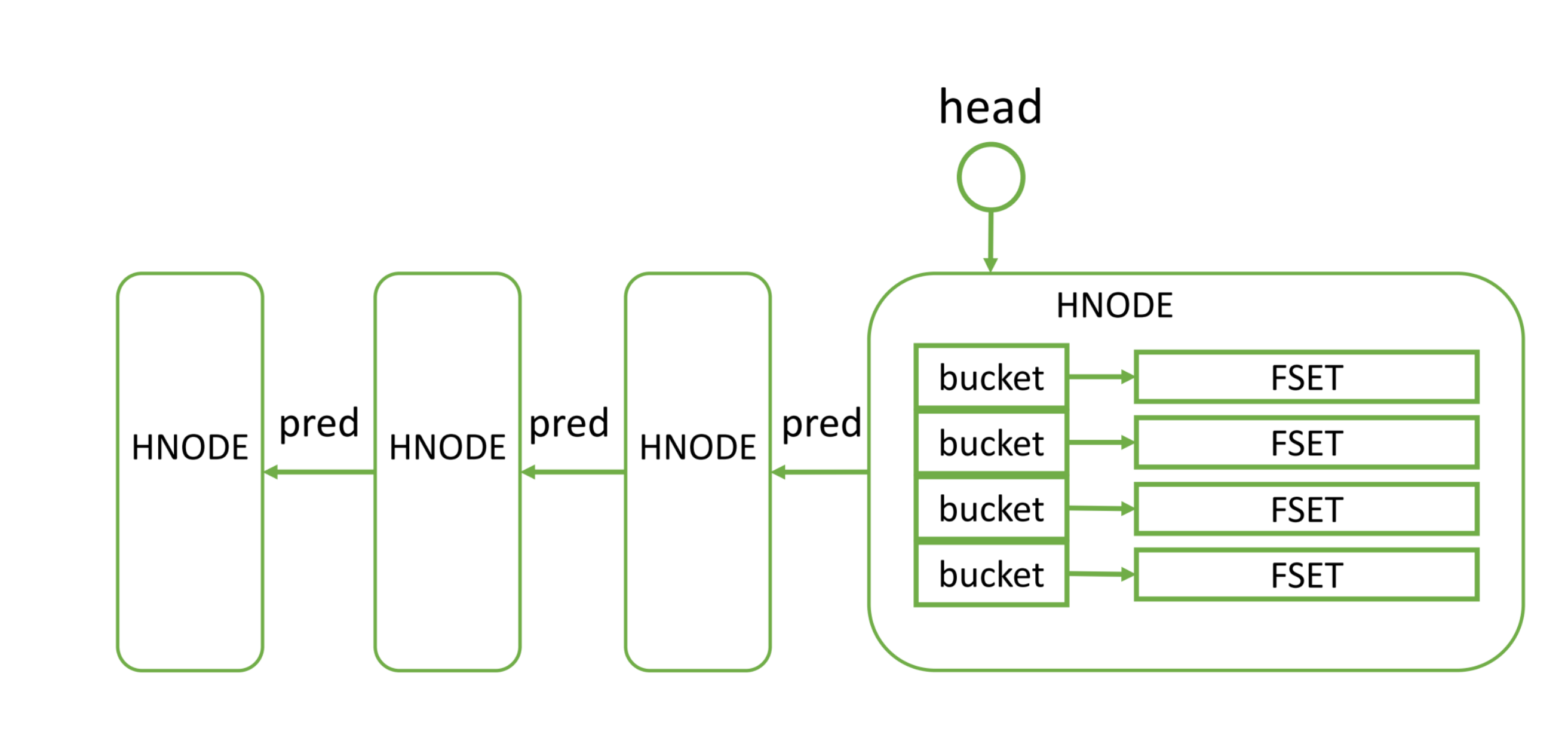}
\caption{The high-level design of the lock-free hash set.}
\label{fig:hashset}
\end{figure}

In applying our framework, keys in each bucket are placed in new node objects, so that buckets contain nodes containing keys. This is to introduce a level of indirection needed for the snap-collector algorithm to work. An iterator traverses the bucket array of the \textit{head} \hnode{},
one bucket at a time. It linearly scans the nodes in each bucket.
If the iterator finds an uninitialized bucket, it follows $pred$ to retrieve the bucket(s) in the previous version.


\subsubsection{Local consistency}

We prove that each atomic step of the hash set's operations is locally consistent. 
There are four atomic steps that can affect an iterator's view. 
The first one is the CAS in \dsins{}(\textit{u}) that replaces a bucket $F$ with an updated copy $F'$ containing $u$.
The change set $R^m$ of this atomic step is $\{u\}$. If an iterator has already read $F$, or is in the process of reading $F$,
then the iterator will not notice the change this CAS makes, so its view will not change. If the iterator has not reached
bucket $F$, it will eventually read the new \fset{} $F'$ containing $u$. All other buckets are unaffected, so in both cases, 
$V(N, T) \Delta V(N, m(T)) \subseteq \{u\}$ for any iterator and node $N$. Hence insert's atomic step is locally consistent. The second atomic step is the CAS in \dsdel{}(\textit{v}) that replaces a bucket $F$ with an updated copy $F'$ omitting $u$.
$V(N, T) \Delta V(N, m(T)) = \varnothing$ in this case, as $u$ has already been marked as logically deleted. 
Therefore it is trivial to prove \del{}'s atomic step is also locally consistent.

The third atomic step is the CAS that a thread makes to initialize a bucket $F$. 
Recall that nodes in $F$ are retrieved from the previous \hnode{}.
If an iterator reaches $F$ before it is initialized, it will traverse the corresponding buckets
in the previous version. The iterator will encounter the same nodes if it reaches $F$
after it has been initialized. Hence this CAS is locally consistent, as it will not change an iterator's view 
($V(N, T) \Delta V(N, m(T)) = \varnothing$). 

The last step is the CAS in a resize that swings \textit{head} to the newly created \hnode{}.
An iterator accessing the buckets of new \hnode{} will find them all uninitialized, so the iterator will read the same nodes regardless of how it is interleaved with the resize CAS.
Therefore the iterator's view is unchanged, meaning this step is locally consistent.

\subsection{Chromatic Tree}
\label{sec:general_technique_for_trees} 
Brown \etal{} \cite{Brown2014} proposed a general technique to implement non-blocking trees 
using the \emph{LLX}, \emph{SCX} and \emph{VLX} primitives, which are multi-word generalizations 
of the load-link (LL), store-conditional (SC) and validate (VL) operations, respectively. They implement the chromatic tree (described in the previous section) using their technique. Recall that updates to the chromatic tree are made in a copy-on-write manner: a connected subgraph is atomically replaced with a new connected subgraph reflecting the update to the tree. These update operations consist of insertion, deletion, and 22 different rotations for rebalancing. 

\subsubsection{Local consistency}
The RB2 operation and its local consistency proof were discussed in the previous section. 
Here, we describe two additional operations, BLK and W1, and prove local consistency. Furthermore, we argue that local consistency proofs of the remaining operations are largely the same. We only consider how these operations change the \emph{structure} of the tree, and not auxiliary information stored in nodes, since iterators are not concerned with the latter. See the full paper by Brown \etal{} \cite{Brown2014} for detailed descriptions and figures of all 22 rotations.

\begin{figure}[H]
\centering
\includegraphics[scale=0.2]{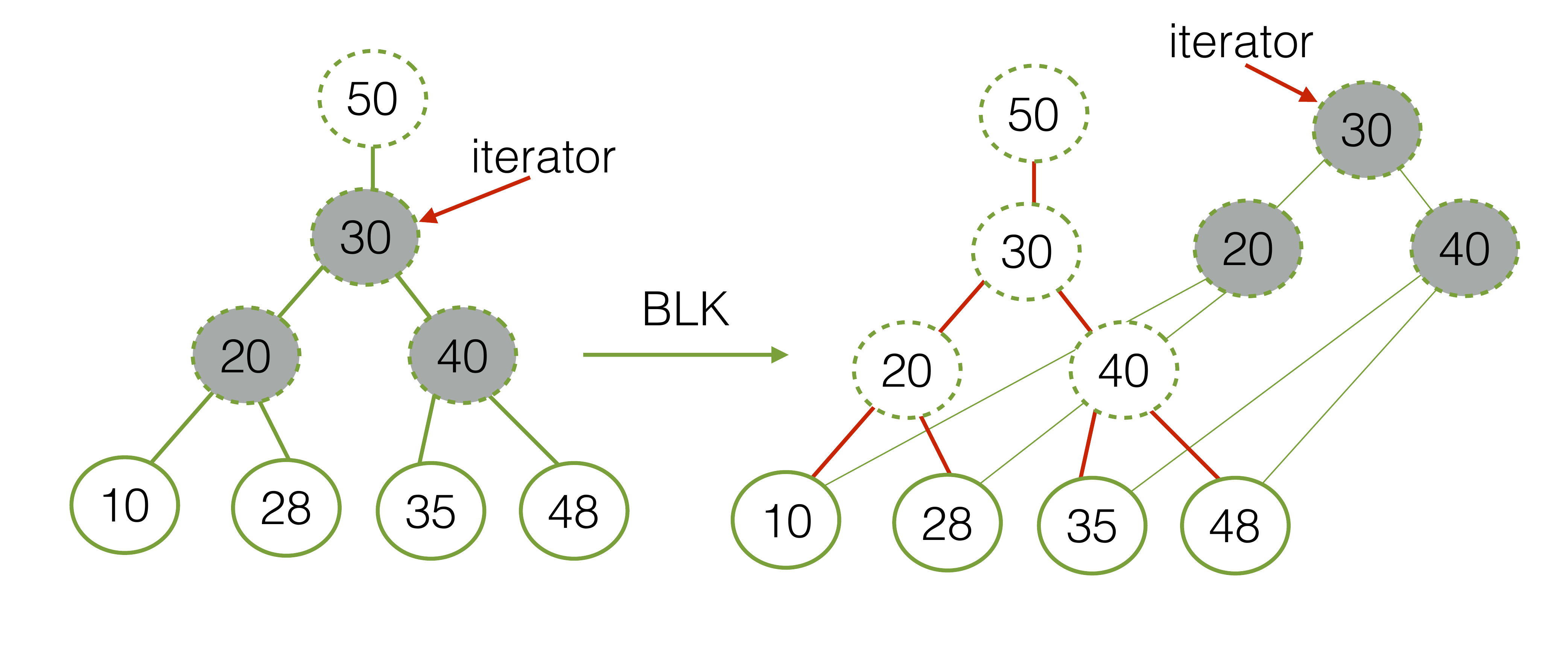}
\caption{The BLK operation.}
\label{fig:blk}
\end{figure}

Since the purpose of each rotation is to help rebalance the tree, and not to insert or delete keys, the change set of each rotation is empty. So in order for a rotation to be locally consistent, it must never change the view of an iterator running concurrently with it. We first consider BLK, illustrated in Figure \ref{fig:blk}. 

As seen in the figure above, BLK replaces a subgraph with an exact copy of it. This is because BLK only updates node weights used for rebalancing; since the chromatic trees nodes are immutable, new copies of the original internal nodes must be made to do this. However, an iterator does not need to know any weight information used by the chromatic tree's set operations, so to an iterator, the tree is essentially unchanged. In particular, the iterator's view will not change after an application of BLK, so BLK is locally consistent. PUSH and W7 also only update node weights while preserving the original structure of the tree, so they are locally consistent as well.

\begin{figure}[H]
\centering
\includegraphics[scale=0.2]{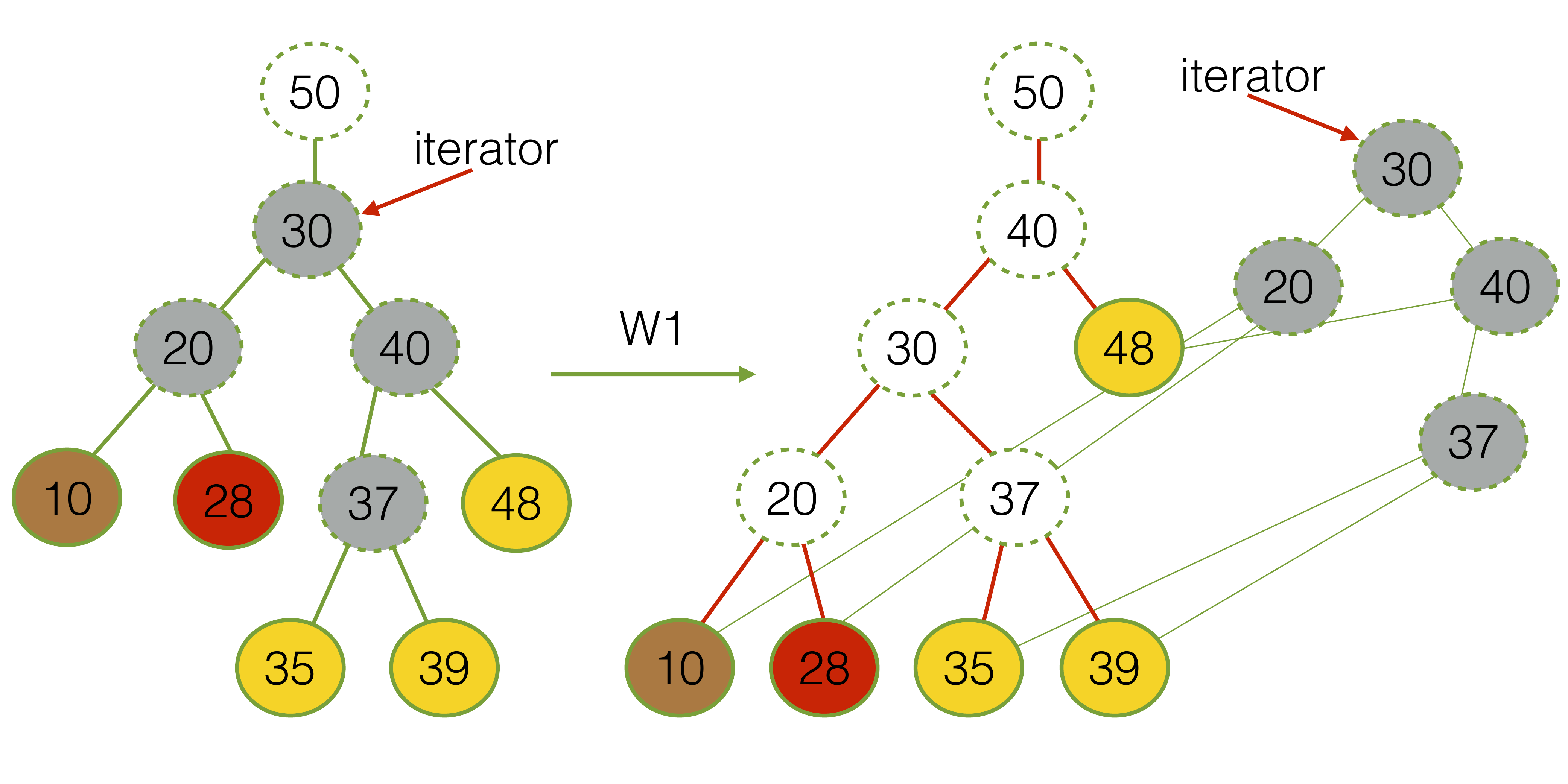}
\caption{The W1 operation.}
\label{fig:w1}
\end{figure}

We now consider W1, illustrated in Figure \ref{fig:w1} above. Unlike BLK, W1 performs non-trivial structural rebalancing of the tree. It is also a more complex rebalance than RB2 (in Figure \ref{fig:RB2}). However, like RB2, the outgoing links of the removed subgraph are not deleted, so an iterator in the middle of traversing a removed portion of the tree will still be able to finish its traversal. So, for example, an iterator at the deleted internal node 30 can reach leaf nodes 35, 39, and 48 via the deleted internal nodes 40 and 37. Since the iterator can still reach all unmodified nodes after W1 is applied, its view is unchanged, so it is locally consistent. This is the same reasoning we used to prove local consistency of RB2. The same reasoning also applies to the remaining rotation operations (RB1, W2, W3, W4, W5, W6). Note that every rotation operation stated in this paper also has an inverse, bringing the count to 22 rotations.

%% file: results.tex


\begin{figure}[t]
    \begin{subfigure}{0.625\textwidth}
        \includegraphics[width=\linewidth]{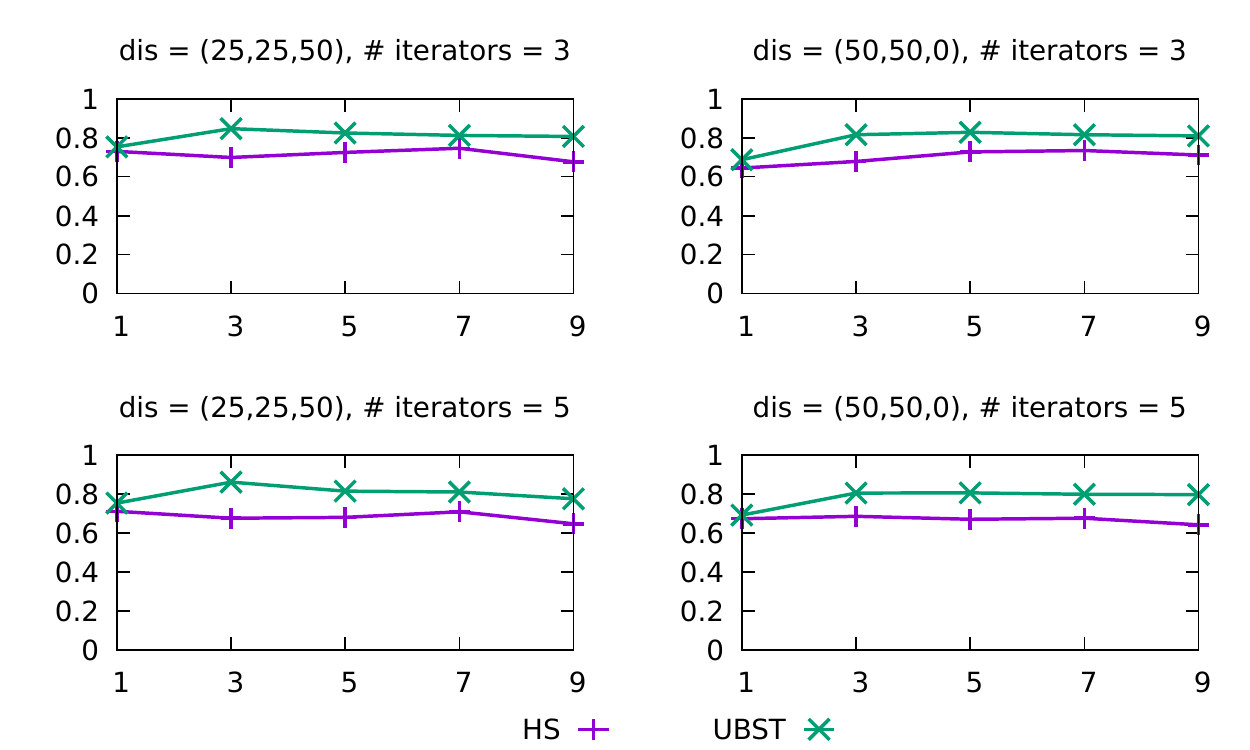}
	\caption{Scalability of the updaters for the key range $[0, 2^{14}]$.}
	\label{fig:scalability_updaters}
    \end{subfigure}%
\hfill
    \begin{subfigure}{0.33\textwidth}
	\includegraphics[width=\linewidth]{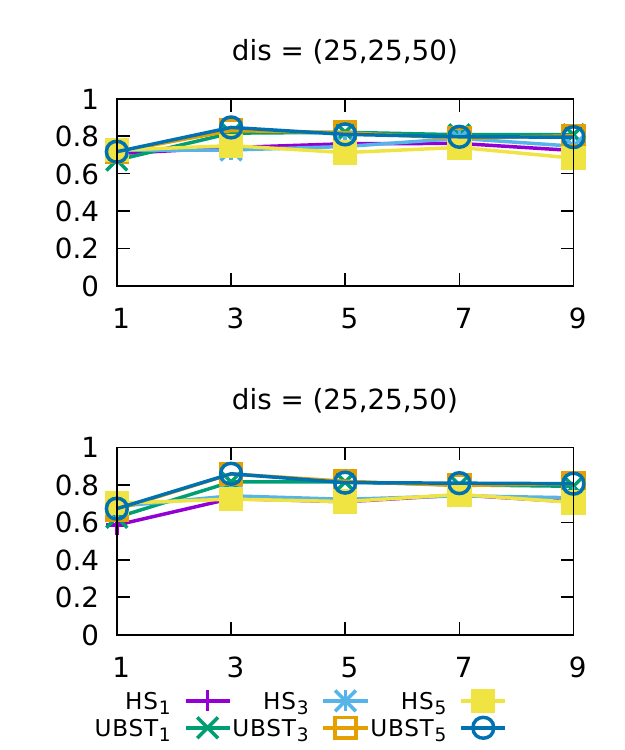}
	\caption{Scalability of the updaters for $[0, 2^{12}]$ key range.}
	\label{fig:scalability_updaters_com}
    \end{subfigure}
    \caption{Updater performance: $x$-axis represents the number of updaters and $y$-axis the slowdown in updater throughput. 
In Figure(b), $\hbox{HS}_i$ and $\hbox{UBST}_i$ are the slowdowns of the hash set and UBST updaters in presence of $i$ iterators. 
}
\end{figure}


\begin{figure}[t]
\centering
\includegraphics[scale=0.4]{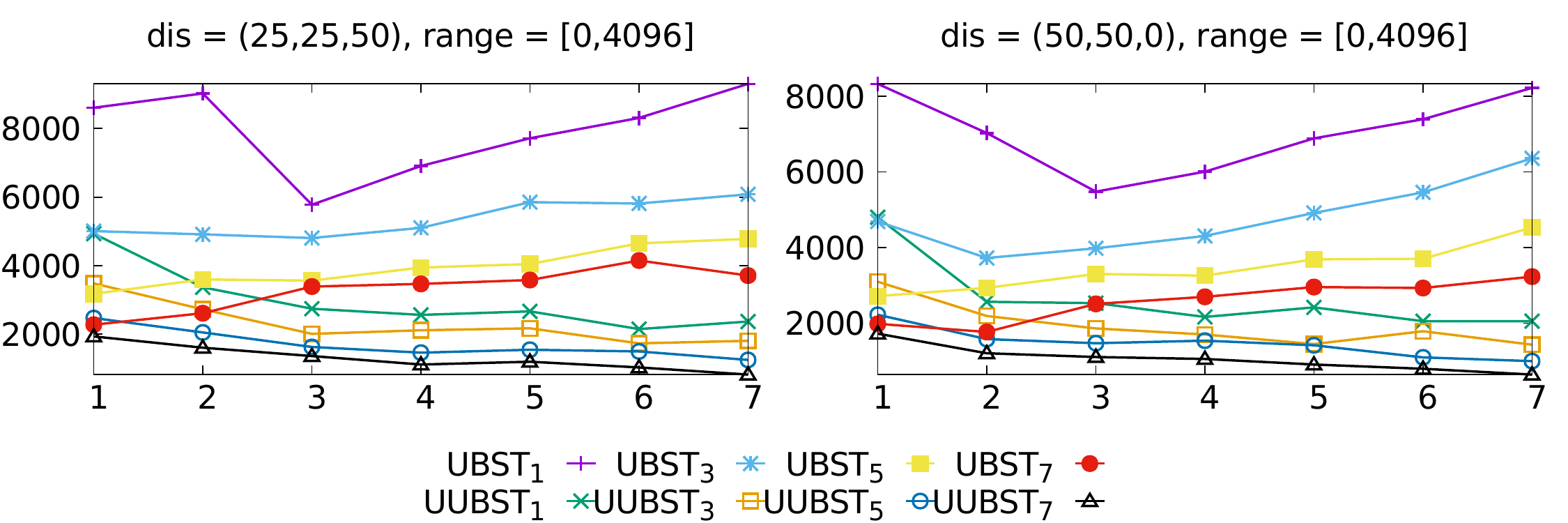}
\caption{Iterator throughput for range $[0, 2^{12}]$: $x$-axis is iterator count and $y$-axis the number of operations by updaters.
$\hbox{UBST}_i$ and $\hbox{UUBST}_i$ are iterator throughput in optimized and unoptimized UBST in presence of $i$ updaters.}
\label{fig:scalability_iterators_com}
\end{figure} 

%
%

We evaluate the performance of the iterator framework applied to an unbalanced binary search tree (UBST) \cite{fast-conc} and a hash set \cite{dyn-hash}. For the hash set, we used the Java code given by the authors, and we implemented UBST in Java according to the pseudocode of the original work. We extended both data structures with an iterator. We ran experiments on Java SE Runtime, version 1.7.0, on a system with 2 Intel(R) Xeon(R) 2.60GHz processors, each processor having 8 cores.

We execute several runs of the following multithreaded benchmark. Some threads are designated as \emph{iterators} which repeatedly call \iter{}. Other threads are designated as \emph{updaters} which call \con{}, \ins{} or \del{}. We consider two distributions of operations: one with 25\% insertions, 25\% deletions and 50\% searches, and the other with 50\% insertions and 50\% deletions. Keys are chosen uniformly at random from each of three ranges: $[0, 2^{12}]$, $[0, 2^{14}]$, $[0, 2^{16}]$. For each distribution and range, we vary the number of updaters between 1 and 9. To maintain a stable performance estimate as the updater count varies, the updaters' operations and the data structure's initial state are predetermined for each distribution and range.

The \emph{throughput} of a run is the total number of update operations completed across a fixed duration. With each run, we measure the \emph{slowdown} in performance of updaters due to iterators as the ratio of the throughput of updaters with no concurrent iterators (called WOI) to the throughput of updaters with iterators (WI). Each experiment consists of phases WOI and WI. Before each phase, the main thread initializes the set (as explained above), then creates iterators and updaters. Threads run for half a second to allow the JVM to warm up, after which each runs 2 seconds. Slowdown is computed after both phases complete.

Results of the experiments are shown in Figure \ref{fig:scalability_updaters}, where each plot shows the average slowdown of updaters over 10 runs for $[0, 2^{14}]$ key range with 3 and 5 concurrent iterators. Overall, we notice that iterators impose an overhead of approximately $20\%$ and $30\%$ on updater throughput of UBST and hash set, respectively, without significantly affecting the underlying set's scalability. The result for UBST is comparable to the results for the linked list \cite{Petrank2013}. Since updates to hash sets are amortized constant time, a small overhead imposed by iterators can account for a large fraction of the total time spent by the updaters. Similar results are obtained as the number of iterators is increased.

Figure \ref{fig:scalability_updaters_com} summarizes experiments for range $[0, 2^{12}]$. Observe that almost all the slowdown factors are approximately 0.8 and 0.7 for UBST and hash set, respectively. We conclude that (i) updaters scale with the number of iterators, and (ii) iterators impose a small overhead on the throughput of the updaters. For the largest range $[0, 2^{16}]$, as the number of iterators increases, scalability is preserved but throughput declines to about 60\%. Since we start with a large initial number ($2^{15}$) of keys, we hypothesize that slowdown is due to the poor cache locality of the huge data structure, and not due to iterators.

Figure \ref{fig:scalability_iterators_com} shows throughput and scalability of iterators with respect to updaters (bottom three lines), with throughput decreasing with more updaters. This is due to iterators merging more reports generated by updaters.

There is a decline in iterations completed per iterator as the number of iterators is increased. This is due to iterators competing to add nodes to the snapshot-list. To reduce contention, iterators are optimized for UBST in the same way Petrank and Timnat \cite{Petrank2013} optimize linked lists: before adding a node, the iterator compares the new node's key with the last key in the snapshot-list, only adding the new key to the snapshot-list if it is greater. This optimization is only possible when keys are traversed in sorted order, e.g. in the tree. 
